**Main Manuscript for**

# Ionization and electron excitation of fullerene molecules in a carbon nanotube. A variable temperature/voltage transmission electron microscopic study

Dongxin Liu,[†] Satori Kowashi,[†] Takayuki Nakamuro,[*,†] Dominik Lungerich,[†,‡, ||] Kaoru Yamanouchi,[†] Koji Harano,[*,†] Eiichi Nakamura[*,†]

[†]Department of Chemistry, The University of Tokyo, 7-3-1 Hongo, Bunkyo-ku, Tokyo 113-0033, Japan
[‡]Center for Nanomedicine (CNM), Institute for Basic Science (IBS), IBS Hall, 50 Yonsei-ro, Seodaemun-gu, Seoul, 03722, South Korea.
[||]Graduate Program of Nano Biomedical Engineering (NanoBME), Advanced Science Institute, Yonsei University, Seoul, 03722, South Korea.
*Correspondence to: Takayuki Nakamuro, Koji Harano, Eiichi Nakamura
**Email:** muro@chem.s.u-tokyo.ac.jp, harano@chem.s.u-tokyo.ac.jp, nakamura@chem.s.u-tokyo.ac.jp

**Author Contributions:** E.N., K.H. and T.N. supervised the project. D.Liu and S.K. carried out the TEM experiments and analyzed the data with T.N., D. Lungerich, K.Y., K.H. and E.N. D. Liu, T.N., K.H. and E.N. cowrote the paper. All authors discussed the results and commented on the manuscript.
**Competing Interest Statement:** Authors declare no competing interests.
**Classification:** PHYSICAL SCIENCE, Chemistry
**Keywords:** excited state chemistry, radiation chemistry, radical cation, fullerene, transmission electron microscopy

**This PDF file includes:**

> Main Text
> Figures 1 to 9
> Tables 1

**Abstract**

There is increasing attention to chemical applications of transmission electron microscopy, which is often plagued by radiation damage. The damage in organic matter predominantly occurs via ionization (radiolysis). Although radiolysis is highly important, previous studies on radiolysis have largely been descriptive and qualitative, lacking in such fundamental information as the product structure, the influence of the energy of the electrons, and the reaction kinetics. We need a chemically well-defined system to obtain such data, and have chosen as a model a variable-temperature and variable-voltage (VT/VV) study of the dimerization of a van der Waals dimer [60]fullerene ($C_{60}$) to $C_{120}$ in a carbon nanotube (CNT) as studied for individual reaction events at atomic resolution. We report here the identification of five reaction pathways that serve as mechanistic models of radiolysis damage. Two of them occur via a radical cation of the specimen generated by specimen ionization, and three involve singlet or triplet excited states of the specimen, as initiated by electron excitation of the CNT followed by energy transfer to the specimen. The pathways were distinguished by the pre-exponential factor and the Arrhenius activation energy. The prototypal reaction path is the radical cation reaction that we saw at <200 K, but, at >350 K, the excited-state reactions dominate. The results illustrate the importance of VT/VV kinetic analysis in the studies of radiation damage, and show that chemical ionization and electron excitation are inseparable but different mechanisms of radiation damage, which has so far been classified loosely under the single term "ionization."

**Significance Statement**

The destruction of specimen molecules by an electron beam (e-beam) is either beneficial as in mass spectrometry capitalizing on ion formation, or deleterious as in electron microscopy. In the latter application, the e-beam not only produces the specimen image but also causes information loss upon prolonged irradiation. However, the atomistic mechanism of such loss has been unclear. Performing single-molecule kinetic analysis of $C_{60}$ dimerization in a carbon nanotube (CNT) under variable-temperature, variable-voltage conditions, we found three reactive species reacting competitively as the voltage and the properties of the CNT were changed. Thus, we identified radical cation, singlet, and triplet excited states, characterized by five different sets of the activation energy and pre-exponential factor. The key enabler of the research was in situ cinematic recording of the whole reaction process, suggesting an upcoming new era of "cinematic chemistry."

**Main Text**

**Introduction**

Since the time of the Knoll/Ruska invention of transmission electron microscopy (TEM) (1), electron microscopy has suffered from information loss during observation often ascribed to the structural changes of a specimen into a different substance, known as radiation damage (2). As summarized recently by Egerton (3), the electron-beam (e-beam) damage in organic matter predominantly occurs via processes triggered by ionization (radiolysis). Although radiolysis is highly important, studies on radiolysis have largely been descriptive and qualitative because of the complexity of the process and the difficulty in quantifying the changes under variable-temperature and variable-voltage (VT/VV) conditions. The first step of the process involves electron-impact ionization (EII), which removes an electron to form a radical cation (RC) of the specimen, or possibly also electron-impact excitation (EIE) where the electron does not fly away to the vacuum but stays in a higher antibonding state in the system. The processes were recently studied in depth for the first time by a thorough quantum chemical study (4). There has been, however, a paucity of experimental mechanistic information; that is, how a specimen is transformed to what product with what level of activation energy and frequency at what acceleration voltage. Single molecules encapsulated in a carbon nanotube (CNT) have often been observed either stably or undergoing well-defined chemical transformations (5,6,7,8,9), and it is primarily because the damage due to secondary electrons is a minimum (10). Drawing an analogous specimen stabilization by a thin metallic coating or deposition on a conductive indium tin oxide substrate in scanning electron microscopy (SEM) (11), we can consider that the electron-rich CNT (Fig. 1*A*) interacts with the specimen molecules and protects them from ionization by filling in the electron vacancy in the RC. In light of the recent characterization of singlet and triplet excitons of CNTs (Fig. 1*A*) (12), we conjectured also that the CNT exciton would excite a molecule in the CNT via energy transfer (ENT) (Fig. 2*A*) (13,14). Electron excitation of graphene under TEM conditions has also been suggested recently (15,16).

We report here the VT/VV kinetic study of the dimerization of a van der Waals (vdW) dimer [60]fullerene ($C_{60}$) to $C_{120}$ in a CNT (17,18,19), in which we found five competing reaction pathways that serve as model pathways of radiolysis damage (Fig. 1*B–F*). We found a marked influence of the temperature, the acceleration voltage, and the properties of the CNT—pristine (prCNT), oxidized (oxCNT), or damaged CNT (dmCNT) (Fig. 2*B*). They have been distinguished by the pre-exponential factor (PEF) and the Arrhenius activation energy ($E_a$). The most frequently occurring reaction was found to occur via singlet ($S_1$) or triplet ($T_1$) species generated by ENT from a CNT exciton (Fig. 1*B,C*) (3). The triplet reaction also occurs when we use an oxidized CNT, which is known to form a triplet exciton (Fig. 1*D*). Electrons of 60 keV in energy cannot energize prCNT to the triplet state, and instead directly ionize $C_{60}$ into RC, which is then reduced by the CNT to a mixture of $S_1$ and $T_1$ states (Fig. 1*E*). This process illustrates how prCNT protects the specimen from radiolysis by "grounding" (Fig. 2*C*) and accounts for the stability of a variety of molecules having low-lying highest occupied molecular orbitals, such as saturated hydrocarbons (20), amides (21), alcohols (22), and inorganic salts encapsulated in a CNT (23,24). We observed the RC when the reaction was performed in a heavily damaged CNT (Fig. 1*F*). The VT/VV behavior of the kinetics agrees with the competitive occurrence of electron excitation and ionization, but not with the atomic displacement damage mechanism, which is the major cause of radiation damage in conductive inorganic materials (25).

**Results**

Light transfers its energy to a zero-dimensional (0D) material via an electric dipole transition (EDT) mechanism with conservation of spin angular momentum (Fig. 1*A*) (26), and the excited species undergo intersystem crossing (ISC) (27) and ENT (28). Being a particle wave, e-beam lacks EDT capability, and causes predominantly plasmon excitation, EII, and atom displacement via momentum transfer (29). EIE of CNT occurs with conservation of spin angular momentum (30). CNT resembles 0D materials due to van Hove singularities (Fig. 1*A*) (31), and we envisaged that the CNT exciton transfers energy to $C_{60}$ encapsulated in CNT (Fig. 2*A*). We monitored the reaction using single-molecule atomic-resolution time-resolved electron microscopic technique (SMART-EM) under VT/VV conditions at 103–493 K and 60–120 keV (10,19).

The thermally forbidden [2 + 2] dimerization of vdW $(C_{60})_2$ does not take place at temperatures <800 K (32). The reaction commences upon photo-irradiation of $C_{60}$ film or solid (33,34), also occurs upon electron irradiation (Fig. 2*A*, box). Further irradiation converts the initially formed [2 + 2] dimer eventually to a short CNT (18) via retro [2 + 2] cycloaddition and a series of Stone–Wales rearrangements (Fig. 3*A*) (35).

In 2011, $C_{60}$ dimerization in CNTs was reported to take place even with 20-keV irradiation, although it then requires a 100-times larger electron dose that at 80 kV (36). The 20-keV energy is far lower than the threshold voltage of carbon atom displacement (CAD, knock-on displacement), and the data suggests a mechanism not going through CAD. A 60-keV electron beam was recently reported to cause reactions of $C_{60}$ sandwiched between two graphene sheets – loss of one carbon atom to form a $C_{118}$ (quasi) dimer via $C_{59}$ (37). This may suggest a difference between a 1D CNT and gapless 2D graphene that lacks van Hove singularities (Fig. 1*A*). This is briefly examined in the present study.

VT-SMART-EM imaging is an emerging experimental tool for the study of kinetics and thermodynamics of individual chemical events (19, 22, 38), and it has provided direct experimental proof of the Rice–Ramsperger–Kassel–Marcus theory (39). In this study, we performed VT-SMART-EM imaging under VV conditions (VT/VV-SMART-EM). Following the reaction conditions developed previously (19), we encapsulated the fullerene molecule in a bundle of pr-, ox-, or dmCNTs, 1.3–1.4 nm in diameter (40), by heating them together at 823 K for 15 h in vacuo. A specimen was deposited on a TEM grid and the time evolution of the [2 + 2] cycloaddition of a $(C_{60})_2$@CNT was visually monitored with a frame rate of two frames per second (s) under e-beam irradiation at 120, 100, 80, and 60 kV, with a constant electron dose rate (EDR) of $3.1 \times 10^5$ e$^-$ nm$^{-2}$ s$^{-1}$ for 120 kV and $5.0 \times 10^6$ e$^-$ nm$^{-2}$ s$^{-1}$ for 60–100 kV throughout this study.

Note that the reaction rate per electron measured in the present work is not affected by the variation of EDR as previously reported (19).

The cycloaddition event was characterized by the change in intermolecular distance from 1.00 nm for the vdW dimer to 0.90 nm for the cycloadduct (Fig. 3*B*), and the $E_a$ and PEF were determined (19). The [2 + 2] cycloadduct features a strained cyclobutane ring in the middle and reverts to two molecules of $C_{60}$ upon heating at >500 K, thus providing compelling chemical evidence that the adduct is the $C_{120}$ cycloadduct (Fig. 3*A*).

**Singlet Dimerization of a $(C_{60})_2$@CNT.** Unlike light that can transfer its entire energy for electron excitation of molecules via an EDT mechanism, the momentum transfer mechanism of an e-beam can transfer its energy to molecules very inefficiently. To assess how inefficient it would be, and to know the consequence of poor efficiency, we performed the kinetic analysis at an acceleration voltage decreasing stepwise from 120 keV to 60 keV.

We first found that a beam of 100–120-keV electrons excites a $(C_{60})_2$ vdW dimer to the $S_1$ state – a pathway expected because the EIE with high-energy e-beam occurs with conservation of spin angular momentum (30). Rather unexpectedly, however, the $S_1$ path was unavailable with a beam of 80-keV electron, which instead opened a $T_1$ path, suggesting that the 80-keV electron generated a triplet CNT exciton. ISC from a singlet exciton of CNT to triplet was documented recently (12). Given the reported $S_0/S_1$ energy difference of ~2.5 eV for $C_{60}$ (41), we estimated that the 100–120-keV e-beam transferred at most ~2.5 eV to $C_{60}$, that is, 2 to $2.5 \times 10^{-5}$ of its kinetic energy (2.5 eV/100–120 keV). Similarly, the $S_0/T_1$ energy difference of ~1.5 eV for $C_{60}$ (41) suggests that at most ~$1.9 \times 10^{-5}$ of the 80 keV was utilized for $C_{60}$ excitation.

Taking these numbers at 80–120 keV into account, we expect that a beam of 60-keV electrons cannot excite the CNT, which typically has a bandgap of 1 eV, and hence we expect no dimerization at 60 keV. Interestingly, excited-state dimerization at 60 keV did take place, albeit very infrequently, suggesting that the excited states are not generated via ENT from the CNT but via the RC (Fig. 1*E*).

In Fig. 4, we summarize all the data of the dimerization at 60–120 keV at 443–493 K. The raw data at 443 K are shown in Fig. 4*A*, where we plot the number of dimerization events observed at intervals of $8.0 \times 10^6$ e$^-$ nm$^{-2}$ irradiation against the total electron dose (TED) up to $3.0 \times 10^8$ e$^-$ nm$^{-2}$ (for 60 s). After in situ monitoring of the reactions of 39–55 $C_{60}$ dimerization events at acceleration voltages of 120 (black), 100 (red), 80 (blue), and 60 kV (green, Fig. 4*A*), we observed three features. First, each reaction event takes place stochastically (42). Second, the occurrence of the events follows the first-order kinetics shown in Fig. 4*B* and *C*, where the $1 - P$ and $\ln(1 - P)$ values are plotted against TED ($P$ = normalized conversion of $C_{60}$). Third, we find three different kinetic profiles.

The rate constants ($k$) at 100 kV are summarized in Fig. 4*D*. The error is arguably large, for several reasons. The CNT is a mixture of entities having different chirality indexes (i.e., the diameters) (43) and, under different physicochemical environments, molecular packing in CNTs changes as the reaction proceeds.

Using the rate constants $k$ obtained at five temperatures, we plotted the Arrhenius plot to obtain the activation energy ($E_a$, slope) and PEF (y-intercept) (Fig. 4*E*). The $E_a$ values at 120 and 100 kV are nearly identical, $33.5 \pm 6.8$ and $32.9 \pm 6.0$ kJ mol$^{-1}$, respectively, hence suggesting the same reaction mechanism. The reaction in CNTs under 120 kV (PEF = $3.9 \times 10^{-4}$ (e$^-$)$^{-1}$ nm$^2$) occurs more frequently than the reaction under 100 kV (PEF = $5.9 \times 10^{-5}$ (e$^-$)$^{-1}$ nm$^2$).

The reaction at 80 kV took place with $E_a$ and PEF values essentially the same as those of the $T_1$ reaction mediated by a triplet-sensitizing oxCNT (see below). We therefore consider the reaction to take place via $T_1$. The $T_1$ species forms less frequently (y-intercept = PEF) but is more reactive than the $S_1$ species (slope = $E_a$). An orbital diagram of the concerted singlet cycloaddition is illustrated in Fig. 4F.

The reaction at 60 kV was markedly slower at ~400 K than the reaction at 100–120 kV (Fig. 4B and Fig. S3). The Arrhenius plot (green, Fig. 4E) deviates from linearity, and we surmise that the kinetics reflects competing $S_1$ and $T_1$, generated by direct EIE and not mediated by the CNT (Fig. 2C; see below). Indeed, the estimated $E_a$ value of 20.8 ± 6.1 kJ mol$^{-1}$ and PEF = 3.7 × 10$^{-7}$ (e$^-$)$^{-1}$ nm$^2$ fall between the values of pure $S_1$ and pure $T_1$.

We estimated the reaction rate of the recently reported dimerization of $C_{60}$ sandwiched between two graphene sheets at 60 kV (Fig. 4E caption) (37), and obtained ln(k) = –21.9. This data placed at 298 K in Fig. 4E (dark green, x) lies close to our 60-kV data.

**Triplet Dimerization of a $(C_{60})_2$@oxCNT.** The oxCNT (Fig. 2B), prepared using KMnO$_4$ oxidation of a CNT (44), has both the π- and σ-carbon skeletons destroyed by chemical oxidation, as demonstrated by infrared (IR) absorption (due to benzophenone-like groups) (45). It is reported to be a triplet sensitizer in solution, as efficient as benzophenone, and has a triplet energy lower than ~2.5 eV (46). We encapsulated $C_{60}$ in oxCNT and studied 30–52 vdW $C_{60}$ dimers (($C_{60})_2$@oxCNT).

The time course of the dimerization events at 120 kV, with a constant EDR of 3.1 × 10$^5$ e$^-$ nm$^{-2}$ s$^{-1}$, is shown in Fig. 5A, and the frequency integrated over time in Fig. 5B. The semilogarithmic plot in Fig. 5C gives the reaction rates at temperatures between 378 and 453 K, and the Arrhenius plot gives the $E_a$ and PEF values (Fig. 5D). The data agree with values obtained for a prCNT at 80 kV (Fig. 4E), suggesting a triplet mechanism (Fig. 5E).

$C_{60}$ dimerization at 103–203 K occurred with induction period (Fig. 6C) (19), during which the π-conjugation of the CNTs was destroyed, as seen in Fig. 6A and B (47). After the induction period, a steady first-order reaction took place. We measured the reaction rate and obtained $E_a$ = 1.7 ± 0.6 kJ mol$^{-1}$ and PEF = 1.3 × 10$^{-7}$ (e$^-$)$^{-1}$ nm$^2$ (Fig. 6D).

The remarkably low $E_a$ value suggests an RC, which is formed by ionization and is expected to be extremely reactive (Fig. 6E). RC formation is expected by the standard damage mechanism of radiolysis (3).

Table 1 summarizes the $E_a$ and PEF data in pr-, ox-, and dmCNTs of the four reaction types (Fig. 1B–E). The kinetic profiles are color coded in black, blue, green, and purple. We consider that the path with $E_a$ values of 32.9–33.5 ± ~6 kJ mol$^{-1}$ in Table 1A (black) took place via $S_1$, first because a high-energy e-beam excites CNT with conservation of spin angular momentum, and second because the values compare favorably (within experimental error) with an $E_a$ value of 28 kJ mol$^{-1}$ (a value calculated from Fig. 5 of ref 48) reported theoretically for $S_1$ [2 + 2] cycloaddition in gas phase (48). We assign the $E_a$ values of 11–15 kJ mol$^{-1}$ in Table 1B as obtained for the oxCNT to the $T_1$ pathway (41) because an oxCNT is an effective triplet sensitizer due to aromatic ketone residues that accelerate relaxation of singlet to triplet (49). The low values of $E_a$ agree with the biradical character of the $T_1$ excited state of $C_{60}$. Similarly, we assign $T_1$ to the 80-keV experiment in a prCNT (Table 1A) because the kinetic data agree with those for an oxCNT in Table 1B. The value of 20.8 kJ mol$^{-1}$ at 60 keV in Table 1A (green) coincides with a value of 23 kJ mol$^{-1}$ determined for photodimerization possibly reflecting ISC from singlet to triplet possibly in a 1:3 ratio (50).

**Discussion**

The SMART-EM study on the electron-impact promoted [2 + 2] cycloaddition mediated by CNTs (Fig. 7) is unique in that we can study in situ the individual reaction events one by one as they take place. The first stage is a fast EIE reaction, characterized by the PEF data. The second stage is a slow thermally driven reaction of excited $C_{60}$ going across an energy barrier with a frequency of $\exp(-E_a/RT)$. We determined the kinetic parameters separately for the two steps by visually monitoring the individual events of the forward cycloaddition of vdW complexes, which excludes the contribution of cycloreversion and reversible collisions from the kinetic data analysis.

The ln(PEF) values represent $\ln(k)$ at $T$ = infinite, and they vary widely between –7.9 and –16.4 (PEF = $3.9 \times 10^{-4} - 1.3 \times 10^{-7}$ $(e^-)^{-1}$ $nm^2$). They are also extremely low in absolute magnitude, indicating that a large number of electrons ($1.0 \times 10^3 - 3.0 \times 10^6$ electrons) are required to form one excited or ionized $C_{60}$ molecule (area of 0.396 $nm^2$) that produces the dimer. The $E_a$ values (slope) reflect the reactivity of these species in the thermal dimerization reaction (Fig. 7, second step).

To describe the efficiency of the reaction, we borrow the concept of external quantum efficiency (EQE) used to evaluate the efficiency of photovoltaic devices—the ratio of the number of electrons and holes generated by a device to the number of incident photons shining on the device from outside. Similarly, we can define the EQE based on the number of dimers formed relative to the TED shining on the CNT. The EQE values of the $S_1$ reaction in a prCNT and the $T_1$ reaction in an oxCNT at 120 kV are $9.8 \times 10^{-4}$ and $2.1 \times 10^{-6}$, respectively, indicating that the latter is nearly 1000 times less efficient because of the infrequent formation of the triplet exciton of the CNT. The very low value of the energy attenuation factor (~$10^{-5}$; from ~100 keV to ~2 eV) reflects the lack of a mechanism for efficient energy transfer from the e-beam to the CNT and the loss of energy to phonon vibration of the CNT and physicochemical processes.

The Arrhenius plots for the four representative reactions in Fig. 8 summarize the present finding. In accordance with the accepted mechanism of radiation damage, the ionization pathway operates in dmCNT (purple). In prCNT encapsulating $C_{60}$ (at >300 K), the ionization is suppressed, and much faster excited-state pathways dominate when the energy of the e-beam is >80 keV (black and blue). When the e-beam energy is 60 keV, it does not excite the CNT and hence $C_{60}$.

The $S_1$ species forms in the reaction of $C_{60}$@prCNT at 120 kV (black) that took place most frequently (the largest ln(PEF) value of –7.9). The other three pathways via EIE or EII occurred ~500 times less frequently. Extremely reactive RC (purple) reacted with near-zero $E_a$ and very small ln(PEF) = –15.9 (51). We estimate the ln(PEF) of CAD of $C_{60}$ to be ~–25 to –27, shown as a gray band in Fig. 8, based on the ln(PEF) of RC and the reported frequency difference of ~$10^5$ between CAD and radiolysis of polymers (52). Because CAD is temperature independent (37), we estimate $\ln(k)$ to be ~–25 to –27. Thus, the carbon loss of $C_{60}$ would occur approximately $10^{-5}$ times more slowly than that of the excited-state reactions. We thus expect CAD to become noticeable only after irradiation with TED of $10^9$–$10^{11}$, a dose 100 times greater than that used for SMART-EM imaging. The probability of the atom displacement depends on the elastic scattering cross section, which decreases as the atomic number decreases.

Putting together the above experimental data and the literature information, we suggest, in Fig. 9, the two mechanistic possibilities of radiolysis/ionization of the specimen (Fig. 9A) and excitation (Fig. 9B,C). They are admittedly incomplete but may serve as a cornerstone for future quantitative studies of radiation chemistry under TEM observation conditions using a high-energy e-beam. Fig. 9A-1 shows ionization of the π-rich $C_{60}$, the standard mechanism of radiation damage (Fig. 1F) (52). We found this path at 103–203 K in a dmCNT, and consider that it also accounts for the CNT damage at low temperatures (cf. Fig. 6A,B). Fig. 9A-2 illustrates $C_{60}$ ionization followed by charge neutralization by prCNT and generation of $S_0$, $S_1$, or $T_1$ $C_{60}$ (Fig.

1$E$). We observed this path at 60 kV (36). In prCNT and with 100–120-keV electrons (Fig. 9$B$), the e-beam generates singlet exciton of CNT, and energy transfer forms S$_1$ C$_{60}$ (Fig. 1$B$). However, the 80-keV electron can only form a less energetic triplet exciton to generate T$_1$ C$_{60}$ (Fig. 1$C$). In Fig. 9$C$, oxCNT generates a triplet exciton and then forms T$_1$ C$_{60}$ (Fig. 1$E$). Note that oxCNT has been known to form triplet excitons, probably because of rapid singlet-to-triplet relaxation.

In summary, the kinetics data summarized in Fig. 8 have shown the importance of VT/VV kinetic analysis in the studies of radiation damage, and show that chemical ionization and electron excitation are inseparable but different mechanisms of the radiation damage, which have often been classified loosely under the single term "ionization." The data also showed that the conducting prCNT with its high-lying filled orbital not only protects the molecule from radiolysis (8) (Fig. 2$C$) but can cause selective chemical reactions if suitable orbital interactions between the molecule and the CNT are available. The complexity of the kinetics of EII and EIE suggests a risk in making any mechanistic interpretations of chemical events seen using TEM without performing VT/VV kinetic analysis. The results also illustrated the potential of "cinematic chemistry," microscopic imaging of dynamic chemical events, for elucidation of the mechanisms of chemical reactions (40).

**Materials and Methods**

**Materials.** Single-walled carbon nanotubes (CNTs, Meijo Arc SO, produced by arc-discharge using Ni and Y catalysts, >99% purity, average diameter 1.4 nm, Lot # 6601316) were purchased from Meijo Nano Carbon Co. Ltd. C$_{60}$ powder (nanom purple ST, >98% purity) was purchased from Frontier Carbon Corporation. TEM grids precoated with a lacey microgrid (RO-C15, for VT experiments; pore size 3–8 μm and carbon thickness 70 nm) were purchased from Okenshoji Co., Ltd. Toluene was purchased from Wako Pure Chemical Industries and purified using a solvent purification system (GlassContour) (53) equipped with columns of activated alumina and supported copper catalyst (Q-5) prior to use. Potassium permanganate was purchased from Tokyo Chemical Industry Co., Ltd and sulfuric acid was purchased from Wako Pure Chemical Industries.

**General.** The water content of the solvent was determined using a Karl Fischer moisture titrator (CA-21, Mitsubishi) to be <10 ppm. Bath sonication for the dispersion of CNTs in toluene was carried out with a Honda Electronics WT-200-M instrument. Oxidative removal of the terminal caps of CNTs was carried out in an electric furnace ASH ARF-30KC. Encapsulation of C$_{60}$ into CNTs was carried out in an electric furnace ASH AMF-20, equipped with a temperature controller AMF-9P. IR spectra were recorded on a JASCO FT/IR-6100 instrument with attenuated total reflection. X-ray photoelectron spectroscopy analysis was carried out on a JPS-9010MC instrument using Mg K$\alpha$ X-rays (1253.6 eV).

**Preparation of samples for SMART-EM.** The C$_{60}$@CNTs prepared above are in solid form and thus difficult to deposit directly on a TEM microgrid. We therefore first dispersed samples in toluene (0.01 mg/mL), in vials, which were then placed in a bath sonicator for 1 h. The aim was to soften the samples so that we could secure intimate contact between the CNTs and the carbon surface of the grid. A 10 μL solution of the dispersion was then deposited on a TEM grid placed on a paper that absorbs excess toluene. The resulting TEM grid was dried in vacuo (60 Pa) for 2 h.

**SMART-EM observation.** Atomic-resolution TEM observations were carried out on a JEOL JEM-ARM200F instrument equipped with an aberration corrector and cold-field emission gun (point resolution 0.10 nm) at acceleration voltages of $E$ = 60, 80, 100, and 120 kV, under 1 × 10$^{-5}$ Pa in the specimen column, and with typical spherical aberration values of 1–3 mm. Calibration of the EDR was conducted following a method described in a previous report (19) $C_{60}$ dimerization at 60–120 kV and $C_{60}$ dimerization in oxCNT were monitored at the temperatures mentioned in the main text and an EDR (the number of electrons per second per nm$^2$) of ca. $3.1 \times 10^5$ e$^-$ nm$^{-2}$ s$^{-1}$ for 120 kV and $5.0 \times 10^6$ e$^-$ nm$^{-2}$ s$^{-1}$ for 60–100 kV at 800,000× magnification. The imaging instrument was a CMOS camera (Gatan OneView, 4,096 × 4,096 pixels), operated in binning 2 mode (output image size 2,048 × 2,048 pixels, pixel resolution 0.20 nm at 1,000,000×). A series of TEM images was recorded every 0.5 s as a superposition of 25 consecutive images of 0.04-s frames (automatically processed on Gatan DigitalMicrograph software) over 5–15 min.

We first surveyed $C_{60}$ encapsulated in CNTs on the screen at 200,000× magnification to identify CNTs for reaction monitoring. Having found bundles of CNTs suitable for kinetic studies, we stopped the beam irradiation and changed the magnification to 800,000×. After waiting for 1 min, until thermal drift of the grid ceased or it was at least relatively relaxed, we commenced observation and movie recording. Focusing was carried out during the collection of images, which was recorded at slightly under-focus conditions (defocus value 10–20 nm). At 80, 100, and 120 kV, we continuously focused on 25–70 molecules in total, with a frame rate of 1.0 s for 5–15 min, until most of the $C_{60}$ molecules oligomerized to form an inner nanotube. At 60 kV, the recording time was set to be 15–20 min, following the results of kinetic studies at 80 kV.

**Temperature control.** The temperatures were controlled by using a heating holder (JEOL EM-21130). The accuracy of the grid temperature was 2–3 degrees (according to the instrument specifications). After the stage temperature was raised to the setting value, we waited at least 30 min before commencing observations, in order to stabilize the stage for minimization of thermal drift.

**Image processing.** The images were collected as a .dm3 or .dm4 format file on Gatan DigitalMicrograph software and processed using ImageJ 1.47t software for .dm3 files (54).

**Visual data analysis for counting reaction events of $C_{60}$ dimerization.** The products of $C_{60}$ dimerization were visually identified following a protocol described in a previous report (19), where molecular structures of [2 + 2] cycloadducts were studied thoroughly using atomic-resolution TEM imaging combined with TEM simulations. The progress of the reactions was studied by analyzing the movies backward, from the end of the reaction, to identify $C_{60}$ dimerization. This procedure eliminates complications due to the intervention of equilibrium caused by thermal cycloreversion. The kinetics of cycloaddition between the fused dimer of $C_{60}$ molecules and $C_{60}$ was excluded from the analysis because the resultant product could possess very different properties.

**Acknowledgments**

We thank Profs Yuriko Ono, Tetsuya Taketsugu, and Riichiro Saito for fruitful discussions on theoretical aspects of the study. This research is supported by Japan Society for the Promotion of Science (JSPS) KAKENHI (JP19H05459, JP20K15123, and JP21H01758). D. Liu thanks JSPS and the Program of Excellence in Photon Science for a predoctoral fellowship. S.K. thanks MEXT (ALPS program). D. Lungerich thanks JSPS, the Alexander von Humboldt Foundation, and the Institute for Basic Science (IBS-R026-Y1) for financial support.

**Figures and Tables**

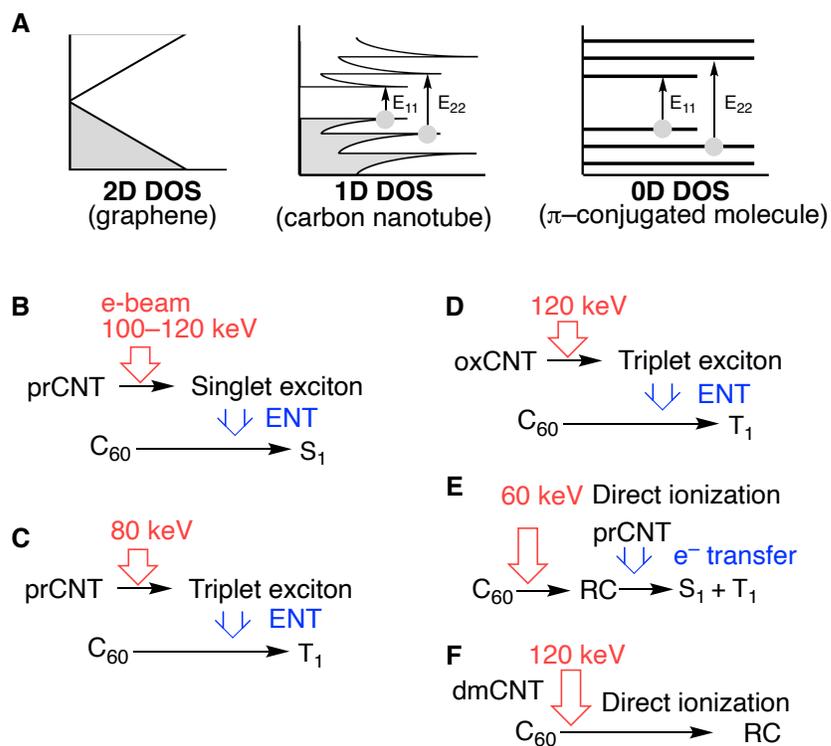

**Fig. 1.** Density of states (DOS) and pathways of $C_{60}$ excitation. (*A*) DOS of 0-D to 2-D materials. (*B–F*) CNT excitation by EIE and ENT from CNT to $C_{60}$.

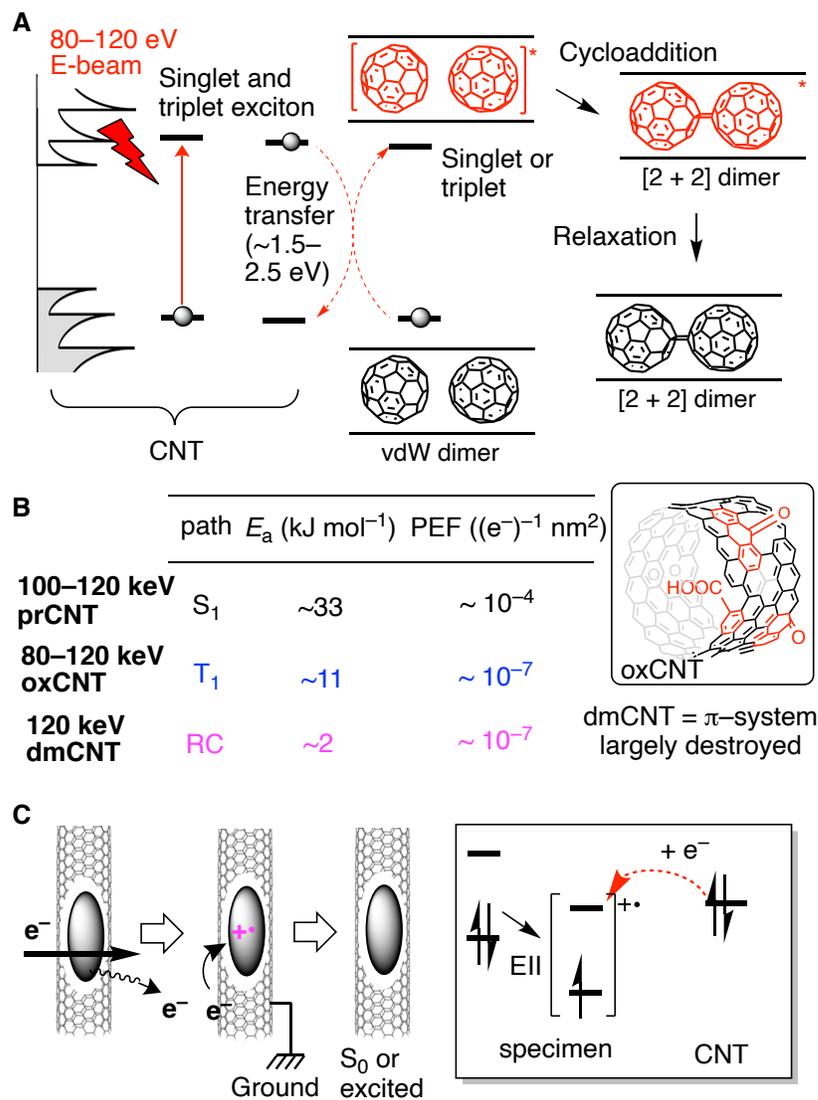

**Fig. 2.** Electron-impact promoted [2 + 2] cycloaddition mediated by CNT–quasi-1D material. (*A*) [2 + 2] Dimerization of $C_{60}$ by EIE of the CNT followed by ENT from the CNT exciton to $C_{60}$. The $E_{22}$ transition is shown as a simplified example of transitions responsible for the $C_{60}$ excitation. (*B*) Representative kinetic parameters of $C_{60}$ dimerization under VT/VV conditions. (*C*) "Grounding" of an ionized specimen molecule by electron transfer from the prCNT.

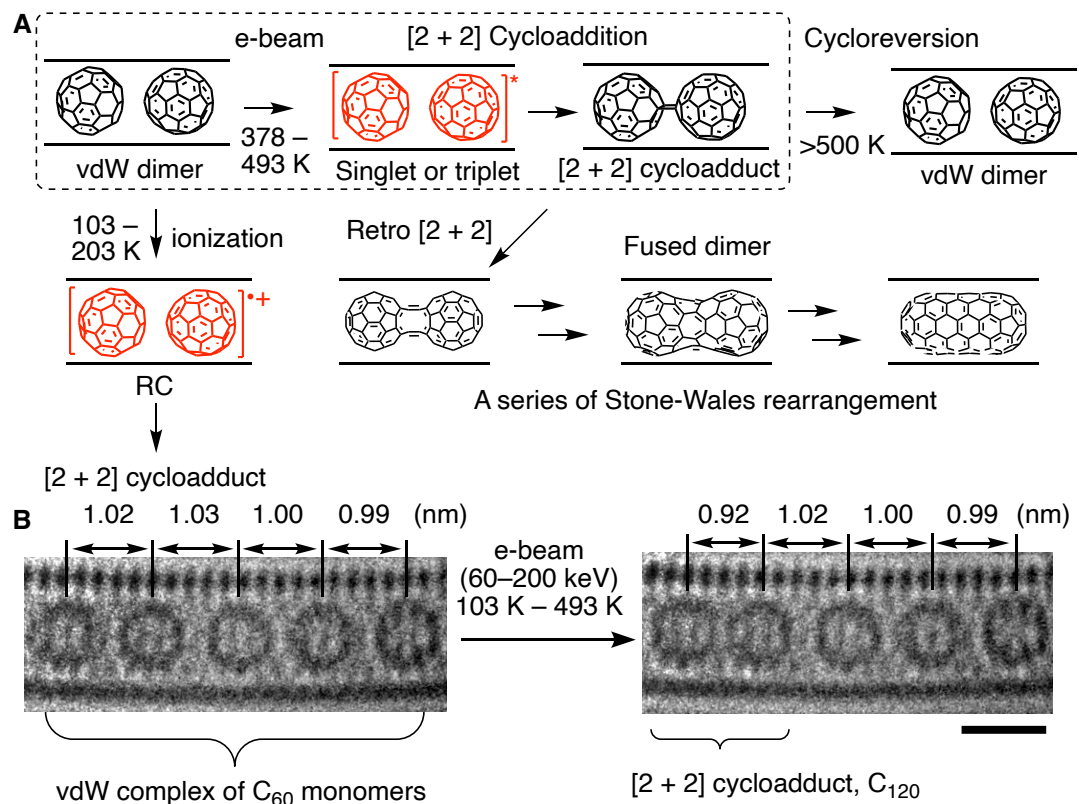

**Fig. 3.** [2 + 2] Cycloaddition via excited state. (*A*) Cycloaddition via excited state and RC as well as retro cycloaddition and further fusion to produce a short CNT. (*B*) TEM images of vdW complexes (intermolecular distance 0.99–1.03 Å) and [2 + 2] dimer (0.92 Å). Fused dimer in (*A*) shows a characteristic intermolecular distance of 0.8 Å. Scale bar = 1 nm.

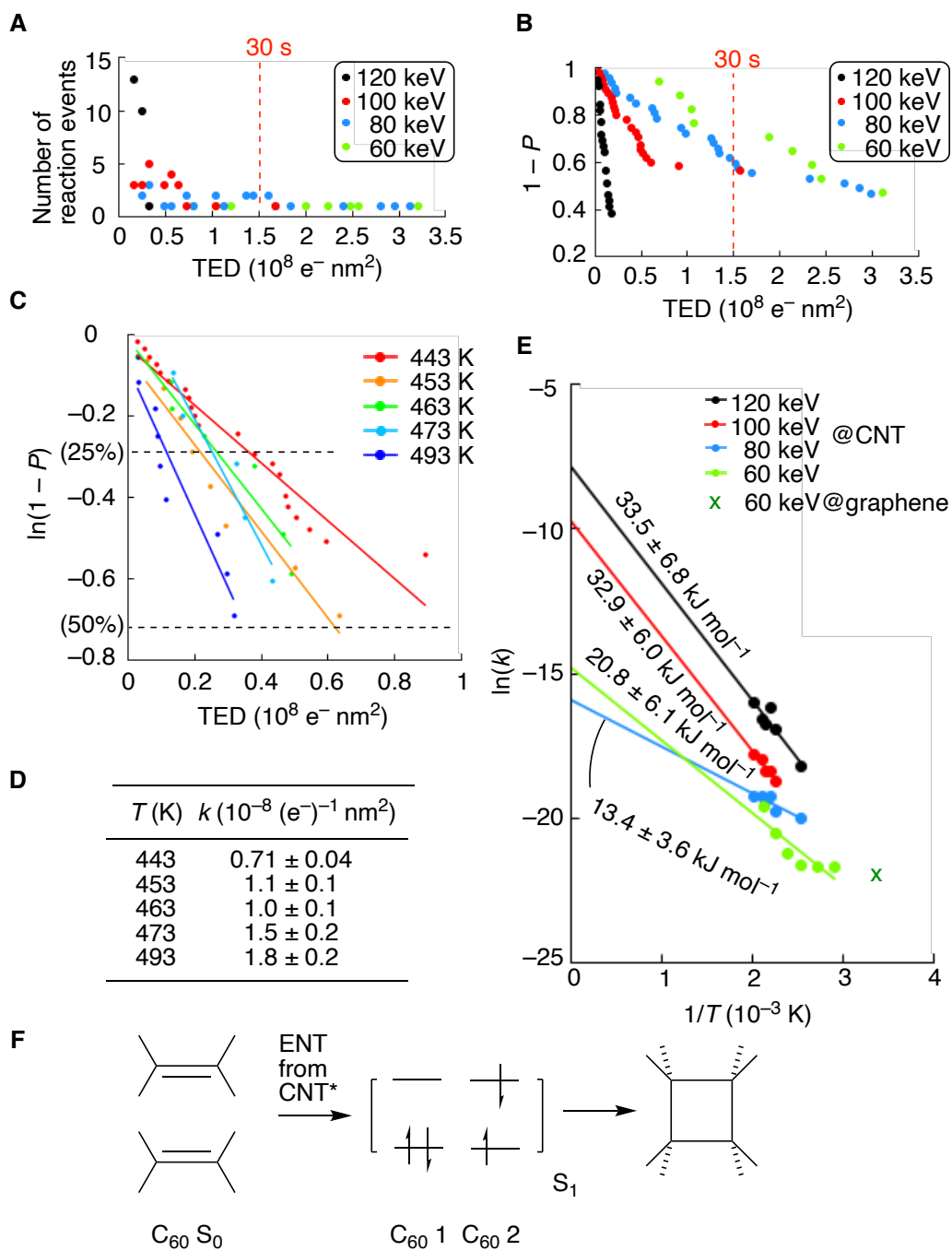

**Fig. 4.** VT/VV-SMART-EM kinetic study of $C_{60}$ dimerization. (*A*) Occurrence of stochastic reaction events of $C_{60}$ dimerization integrated over every $8.0 \times 10^6$ e$^-$ nm$^{-2}$ at 443 K plotted against TED (Table S3, Fig. S2). (*B*) Reaction progress of $C_{60}$ dimerization at 443 K. (*C*) Semilogarithmic plot of $C_{60}$ dimerization at 100 kV above 443 K and first-order kinetic fitting shown as solid lines. (*D*) Reaction rate constants of $C_{60}$ dimerization at 100 kV obtained via linear fitting of (*C*). (*E*) Arrhenius plot of $C_{60}$ dimerization. The green plot is for the 60-keV reaction, where the slope at higher temperature ($1/T = 2$ to $2.4 \times 10^{-3}$) is close to that of the $S_1$ path (black, red) and that at lower temperature (2.5 to $3 \times 10^{-3}$) is close to the $T_1$ path (blue). The x indicates the ln($k$) value

for dimerization of $C_{60}$ sandwiched between graphene sheets estimated from Fig. 4 in ref 37. (*F*) Mechanistic sketch of the $S_1$ cycloaddition.

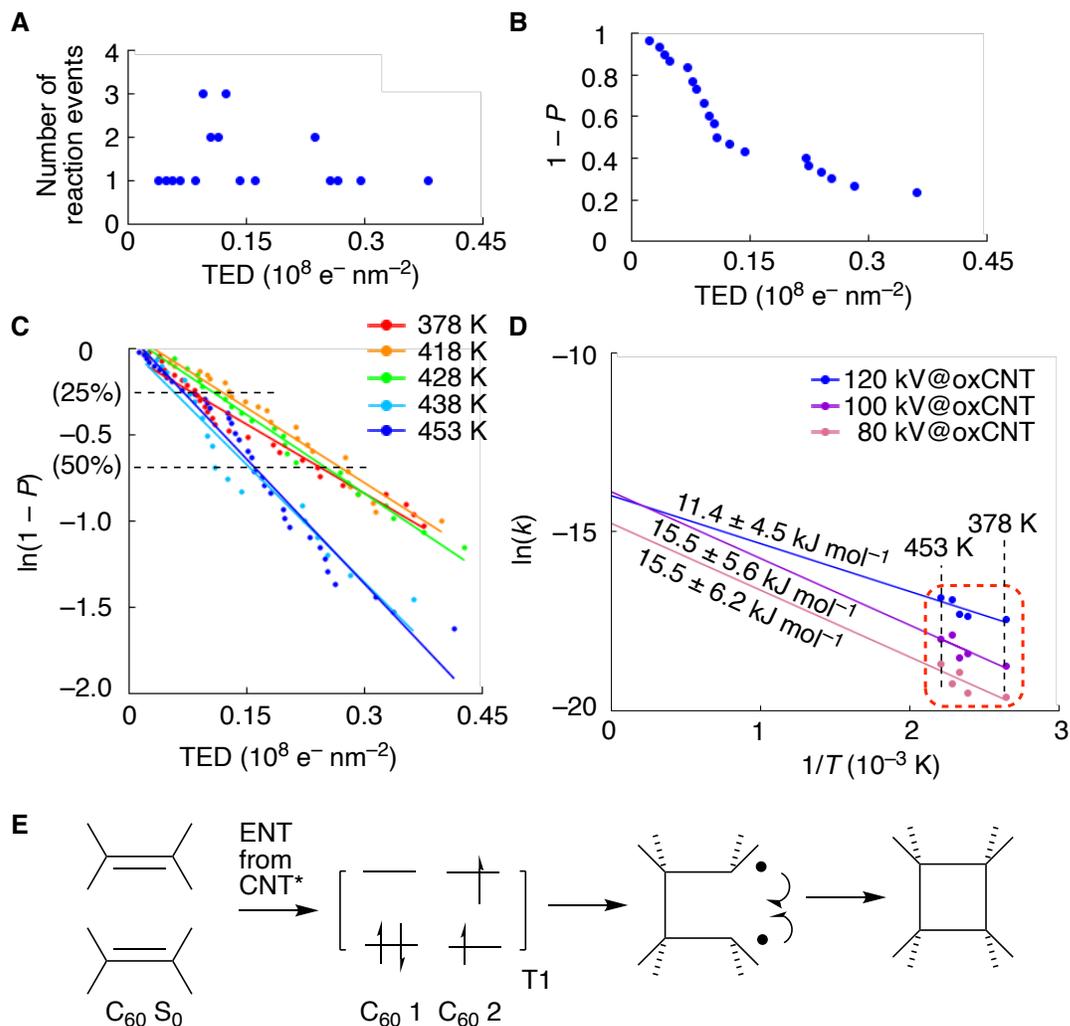

**Fig. 5.** Kinetic study of $C_{60}$ dimerization in an oxCNT. (*A*) Occurrence of stochastic reaction events of $C_{60}$ dimerization inside oxCNTs at 120 kV integrated over every $8.0 \times 10^6$ e$^-$ nm$^{-2}$ at 438 K for a $(C_{60})_2$@oxCNT plotted against TED (Table S4, Fig. S4-5). (*B*) Reaction progress of $C_{60}$ dimerization inside oxCNTs at 120 kV. (*C*) First-order kinetics of $C_{60}$ dimerization inside oxCNTs at 120 kV. (*D*) Arrhenius plot of $C_{60}$ dimerization inside oxCNTs at 80–120 kV. (*E*) Mechanistic sketch of the $T_1$ reaction.

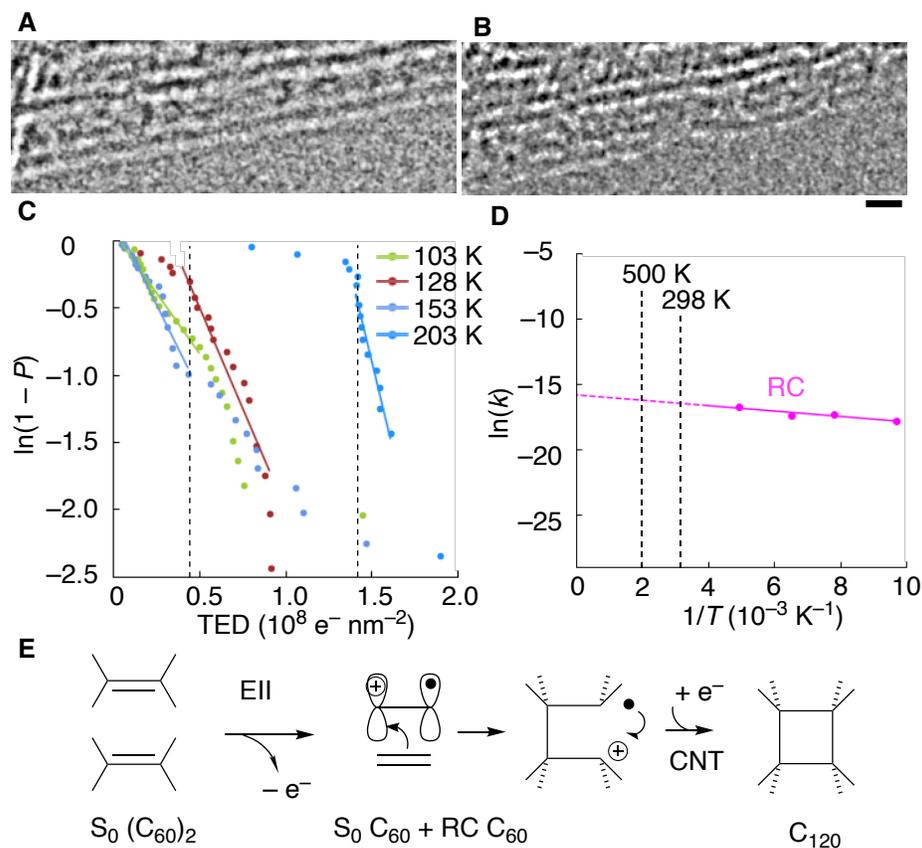

**Fig. 6.** Dimerization at 103–203 K via RC. (*A,B*) The $C_{60}$@prCNT decomposes after prolonged irradiation at 153 K to produce $C_{60}$@dmCNT. Scale bar = 1 nm. (*C*) First-order kinetics of $C_{60}$ dimerization in dmCNTs at 120 kV. Dotted lines show the end of induction period at 128 K and 203 K, from where the rate was calculated. (*D*) Arrhenius plot of $C_{60}$ dimerization in a dmCNT at 120 kV. (*E*) Mechanistic sketch of the RC reaction.

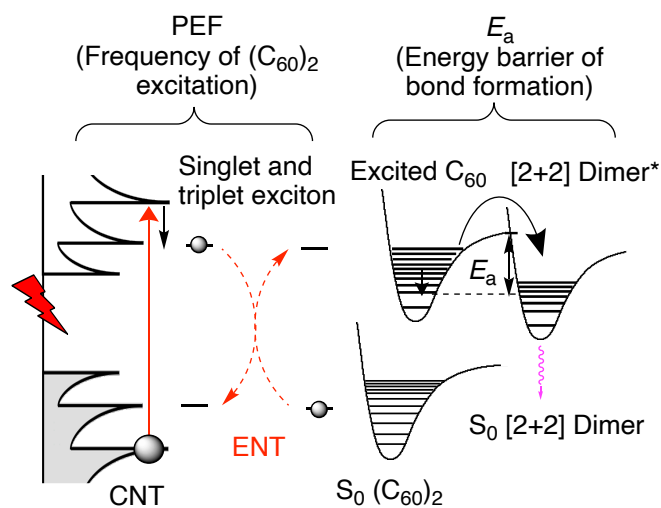

**Fig. 7.** PEF and $E_a$, representing EIE/ENT and cycloaddition, respectively. Of two possible mechanisms of ENT, the Förster mechanism of ENT is shown. The $E_{22}$ transition is shown as an example of transitions responsible for $C_{60}$ excitation. The $E_{33}$ transition followed by thermal relaxation is shown as an example of the processes involved in the $C_{60}$ excitation.

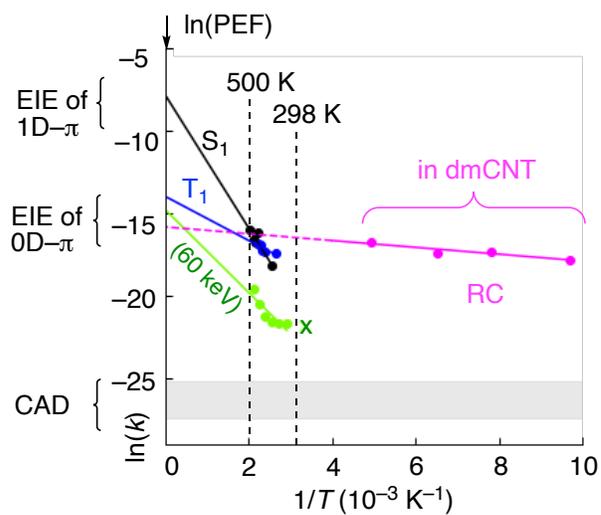

**Fig. 8.** Arrhenius plot for four representative reactions. Black: The 120-kV data in Table 1*A* via $S_1$. Blue: the 120-kV data in Table 1*B* via $T_1$. Green: the 60-kV data in Table 1*A* via direct EIE. Purple: the 120-kV data in Table 1*C* via RC. Gray band: a range of ln($k$) values for temperature-independent CAD estimated from ln(PEF) of radiolysis. The x indicates the ln($k$) value for dimerization of $C_{60}$ sandwiched between graphene sheets estimated from Fig. 4 in ref 37.

## A Direct ionization of $C_{60}$

### A–1 Direct ionization generates RC of $C_{60}$ (Fig. 1F)

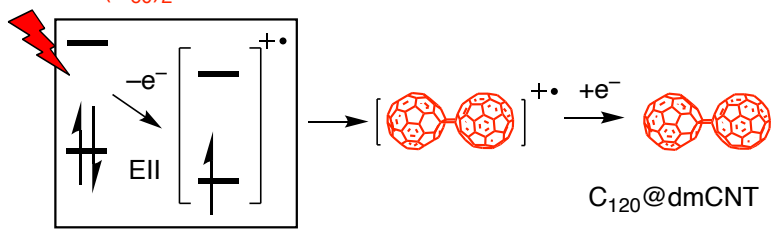

### A–2 Ionization/charge neutralization by prCNT (Fig. 1D)

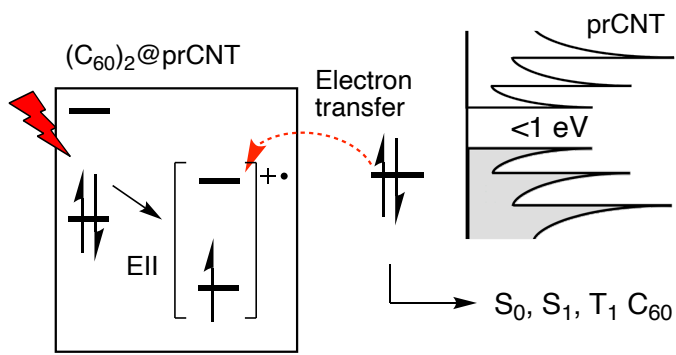

## B Singlet exciton of prCNT (Fig. 1B)

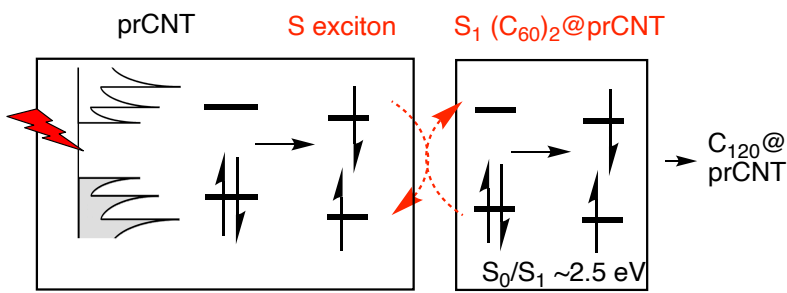

## C Triplet exciton of oxCNT (Fig. 1D)

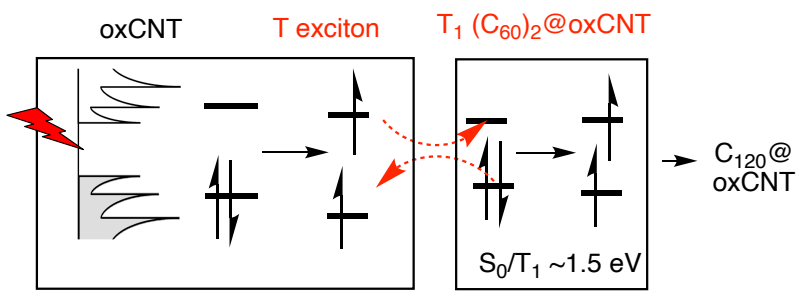

**Fig. 9.** Four pathways available for activation of vdW $(C_{60})_2$@CNT for [2 + 2] cycloaddition. (*A*) Two paths following the initial ionization. (*A*-1) Ionization of $C_{60}$ to generate radical cation at 103–203 K in a dmCNT. (*A*-2) Ionization of $C_{60}$ followed by charge neutralization to generate an excited state taking place with a 60-kV e-beam. (*B*) Singlet CNT exciton generates $S_1$ $C_{60}$ with a 100–120-kV e-beam. (*C*) Triplet oxCNT generates $T_1$ $C_{60}$ with an 80-kV e-beam.

**Table 1.** $E_a$ and PEF values obtained from the Arrhenius plot of the $C_{60}$ dimerization events: (*A*) $C_{60}$ dimerization in a prCNT, (*B*) in a oxCNT, and (*C*) in a dmCNT. Color coding according to the reactive species.

**A** $(C_{60})_2$@prCNT

| e-beam (keV) | $E_a$ (kJ mol$^{-1}$) | ln(PEF) | PEF ((e$^-$)$^{-1}$ nm$^2$) | path |
|---|---|---|---|---|
| 120 | 33.5 ± 6.8 | −7.9 ± 1.8 | 3.9 × 10$^{-4}$ | $S_1$ |
| 100 | 32.9 ± 6.0 | −9.7 ± 1.6 | 5.9 × 10$^{-5}$ | $S_1$ |
| 80 | 13.4 ± 3.6 | −15.9 ± 1.0 | 1.2 × 10$^{-7}$ | $T_1$ |
| 60 | 20.8 ± 6.1 | −14.8 ± 1.8 | 3.7 × 10$^{-7}$ | $S_1/T_1$ |

**B** $(C_{60})_2$@oxCNT

| e-beam (keV) | $E_a$ (kJ mol$^{-1}$) | ln(PEF) | PEF ((e$^-$)$^{-1}$ nm$^2$) | path |
|---|---|---|---|---|
| 120 | 11.4 ± 4.5 | −14.0 ± 1.3 | 8.3 × 10$^{-7}$ | $T_1$ |
| 100 | 15.5 ± 5.6 | −13.9 ± 1.9 | 9.3 × 10$^{-7}$ | $T_1$ |
| 80 | 15.5 ± 6.2 | −14.8 ± 1.8 | 3.9 × 10$^{-7}$ | $T_1$ |

**C** $(C_{60})_2$@dmCNT

| e-beam (keV) | $E_a$ (kJ mol$^{-1}$) | ln(PEF) | PEF ((e$^-$)$^{-1}$ nm$^2$) | path |
|---|---|---|---|---|
| 120 | 1.7 ± 0.6 | −15.9 ± 0.5 | 1.3 × 10$^{-7}$ | RC |